\documentclass[preprint,showpacs,pre]{revtex4}
\usepackage{graphicx}
\usepackage{dcolumn}
\usepackage{bm}
\begin{document}
\title{Fractal dimensions of a green broccoli and a white cauliflower}
\author{Sang-Hoon Kim}
\email{shkim@mail.mmu.ac.kr}
\affiliation{Division of Liberal Arts,
Mokpo National Maritime University, Mokpo 530-729, {\rm and}
 Institute for Condensed Matter Theory, Chonnam National University,
Gwangju 500-757, Korea}
\date{\today}
\begin{abstract}
The fractal structures of a green broccoli and
 a white cauliflower are investigated by box-counting
 method of their cross-sections.
The capacity dimensions of the cross-sections are
$1.78 \pm 0.02$ for a green broccoli and 
  $1.88 \pm 0.02$ for a white cauliflower,
and both are independent of their directions.
From the results we predict that the capacity
dimensions of the bulks are about $2.7$ for
a green broccoli and $2.8.$ for a white cauliflower.
 The vertical cross-sections of the two plants
 are modeled into self-similar sets of triangular
 and  rectangular trees.
 We discuss the conditions of the fractal
objects in the model trees.
\end{abstract}
\maketitle

A green broccoli(GB) and a white cauliflower(WC)
are varieties of vegetables 
 grown for an edible immature flower panicles
or  head of condensed flowers and flower stems.
They are some of the most broadly nutritious of all 
common vegetables, and, nowadays,
  known as nice  foods for healthy diet.
More interestingly, in the view of physical science
the terminal clusters of the vegetables 
 has been known as typical fractals among
living organisms\cite{man}.

Grey and Kjems discussed the fractal structure of a cauliflower,
 and they suggested  that the fractal dimension of it
could be larger than that of a broccoli\cite{grey}.
Later, the cross-section of a broccoli in a vertical direction
 has been modeled into a self-similar set of a
Pythagoras tree\cite{pei}.
Recently, Romera et. al. suggested a mathematical model of the 
horizontal cross-section of a cauliflower as
a sequence of a baby Mandelbrot set\cite{rom}.
It is clear that the dimensionality is the most
fundamental concept of fractal analysis.
 Nevertheless, surprisingly,
in spite of those excellent articles, 
 the fractal dimensions of GB and WC are not known yet.

In this article, we measure the fractal dimensions of the
cross-sections of GB and WC by direct scanning method, and then 
predict those  of the bulks theoretically.
Next, we create mathematical models for the cross-sections
of GB and WC, and compare those with the experimental results.
The condition of creating the fractals
in the mathematical models is discussed, too.
 
 The basic concept of dimensions
in our model has the property as follows:
 Let us introduce that  $\delta$ is a measurement scale 
and that the measurement is $M_{\delta}(F)$. Then, the fractal
dimension $D$ of a set $F$ is determined by the power law: 
$M_{\delta}(F) \sim \delta^{-D}$.
If $D$ is a constant as $\delta \rightarrow 0$, 
$F$ has a dimension of $D$\cite{sch}. 
The term ``capacity dimension" or
``box-counting dimension" is most widely used, 
because it can be easily applied to
fractal objects. It is defined as
\begin{equation}
D =\frac{\log N_{\delta}(F)}{-\log \delta},
\label{20}
\end{equation}
where $N_{\delta}(F)$ is the smallest number of sets of diameter
at most $\delta$ which can cover $F$\cite{fal1,fal2}.

Fractal object embedded in one- or two-dimension is easy to
measure relatively compared to other higher dimensions.
However, fortunately, the dimensions of the cross-sections
 are generally known to be related with those of the bulk. 
 We can propose an  {\it ansaz}: Let $D_c$ be a fractal dimension of a
cross-section embedded in two dimensions. Then, that
of the bulk embedded in three dimensions can be written as
\begin{equation}
D = \frac{3}{2N}\sum_{i=1}^{N} D_{c_i},
\label{30}
\end{equation}
where $i$ represents every possible cross-section.
If the fractal dimension of the cross-section is independent of 
the directions, Eq. (\ref{30}) is simply written as $D=(3/2)D_c$.

We prepared several GB and WC about $300g$ in mass 
and $15cm$ in diameter. We cut half of them in horizontal
direction, and the other half in vertical direction.
Then, we scanned the cross-sections by a scanner
and read the images into black and white.
As we see in Fig. 1, the two figures by the two perpendicular
directions look totally different.

Next, we read the images into matrix of numbers.
The numbers are the reflectivity of each pixels.
The black backgrounds produces 0's, and 
the white images produce large numbers of
the order of $10^5.$ 
Let us call the non-zero number 1 for convenience.
The size of the matrix, $M_1 \times M_2$, is about
$10^3 \times 10^3$.
Note that the size of a pixel is an order of 100$\mu m$.

In order to measure the fractal dimension of the cross-sections,
 we count the non-zero numbers by the box-counting method.
It becomes the smallest number of sets to cover the white images
by $100 \mu m \times 100 \mu m$.
In a half reduction procedure of the matrix, 
a $4 \times 4$ component becomes 1
 if it contains any non-zero number, 
otherwise it becomes 0. 
Then, the number of 1's becomes the smallest number of
sets to cover the white images by $200 \mu m \times 200 \mu m$.
And so on.
In the reducing steps in half, that is
$M_1/2^n \times M_2/2^n, n=1,2,3...$,
the conversion(reverse transformation)
 from the reduced matrices to images
is plotted in Fig. 1 for the first three steps.
We see that the basic structure of the cross-sections
remains unchanged in the procedure.

Fig. 2 is the log-log plot for the horizontal direction (a)
and the vertical direction (b).
The slopes are the capacity dimensions of the cross-sections 
of GB and WC.
Repeating these procedures for other GB and WC,
we observed  similar values of  $D_c$ as
 $1.78 \pm 0.02$ for GB and  $1.88 \pm 0.02$ for WC.
Surprisingly, the two slopes for different directions are
very similar.
Therefore, from Eq.\ (\ref{30}) we predict that the capacity
dimensions are about 2.7 for GB and 2.8 for WC.

The modeling of nature should be as simple as it can,
and at the same time, it should contain the basic structure of
the nature. 
Furthermore, if some physical quantities of the model can
match the real systems, it will guarantee more credits to the
mathematical model.
We modeled the cross-sections of GB and WC in vertical direction as
a triangular tree(TT) or a rectangular tree(RT) in Fig. 3.
TT has two equilateral sides and RT has three equilateral sides.
 These are the simplest models of a self-similar set
 that has a single scale factor $t$ or $r$.
The scale factors $t$ of TT and $r$ of RT
  have the relations with the angles $\theta$ and $\phi$ in Fig. 3. 
\begin{eqnarray}
t&=&\frac{1}{2\cos\theta},
\nonumber \\ 
r&=&\frac{1}{2\cos \phi + 1}.
\label{40}
\end{eqnarray}

Downsizing the trees by the ratio $t$ or $r$,
it creates two or three branches.
Therefore, the fractal dimensions of TT and RT are
\begin{eqnarray}
D^{(t)}_c &=&\frac{\log 2}{-\log t},
\nonumber \\
D^{(r)}_c &=&\frac{\log 3}{-\log r}.
\label{50}
\end{eqnarray}
The areas of the TT and RT have the following series:
$1, 2 t^2, 4 t^4, 8 t^6,... (2 t^2)^n,...$ and
$1, 3 r^2, 9 r^4, 27 r^6,... (3 r^2)^n,...$
where $n=0, 1, 2, 3, ...$.
Because the fractal is
the limit of the series, it should converge as $n$ increases.
Therefore, the conditions of convergence of the series are
$2 t^2 <  1$ and  $3 r^2 <  1$.
It corresponds to
$\theta < 45^\circ$ and $\phi < 68.5^\circ$
by Eq.\ (\ref{40}).
Note that it is clear from $D_c < 2$ in Eq. (\ref{50}).

For example, if we match the TT with GB and RT with WC,
we obtain the following  relations to create fractals;
Since the $D_c$ of GB is 1.78, 
the scale factor $t=0.68$ and the angle $\theta=43^\circ$
from Eqs. (\ref{40}) and (\ref{50}).
By the same method, if the $D_c$ of the WC is 1.88, 
then  $r=0.56$ and  $\phi=67^\circ$.
As the angle $\theta$ approaches $45^\circ$ or
$\phi$ approaches $68.5^\circ$, the dimensions of the trees
go to {\it two} because the limit bents to cover the two dimensional
surface. On the other hand, as the angle $\theta$ or 
$\phi$ approaches $0^\circ$,
the dimension of the tree goes to {\it one}
 because the limit goes to a long rod.

We measured the fractal dimensions of the cross-sections
 of a green broccoli and white
cauliflower by a direct scanning method.
They were $1.78 \pm 0.02$ and  $1.88 \pm 0.02$, and almost
independent of the directions of the cross-sections.
 From these results we predict that
 the fractal dimensions of the bulks 
 are about 2.7 for GB and  2.8 for WC.

We created  mathematical models for the vertical directions
with only one scale factor.
It is a triangular or a rectangular tree of three or four
equilateral sides.
We suggested the condition of creating fractals in our model,
and matched the models into experimental measurements.
This method of the scanning cross-sections 
and mathematical models from  polygonal trees
can be widely applied to many complex
bulk fractals in nature.

\newpage

 \begin{figure}
 \caption { The scanned image, top left,  and the images
 from half-reduction procedure and reverse transformation.
From the top left and clockwise direction,
$M_1 \times M_2$, $M_1/2 \times M_2/2$,
 $ M_1/4 \times M_2/4$, and $M_1/8 \times M_2/8$.
(a) Horizontal images of a green broccoli.
(b) Vertical images of a green broccoli.
(c) Horizontal images of a white cauliflower.
(d) Vertical images of a white cauliflower.}
 \caption { The log-log plots of the two perpendicular directions.
 The slopes are $D_{c}'s$
(a) a green broccoli, (b) a white cauliflower }
\caption {The self-similar sets with single scale factors each.
(a) Triangular tree(TT), (b) Rectangular tree(RT).}
\end{figure}
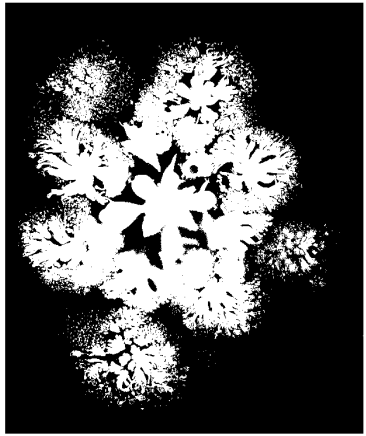
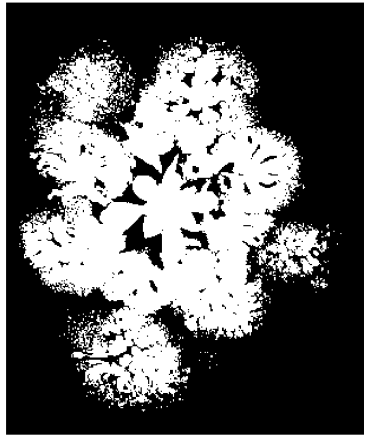
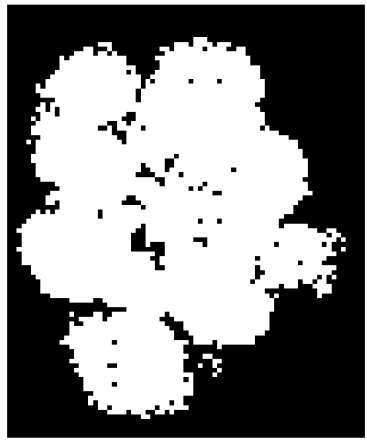
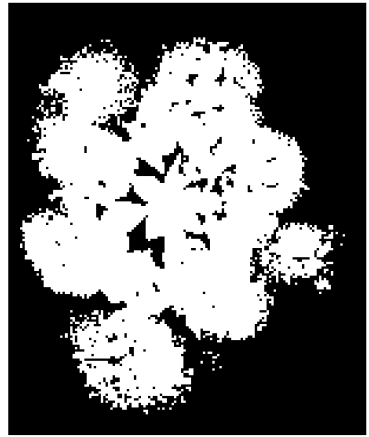
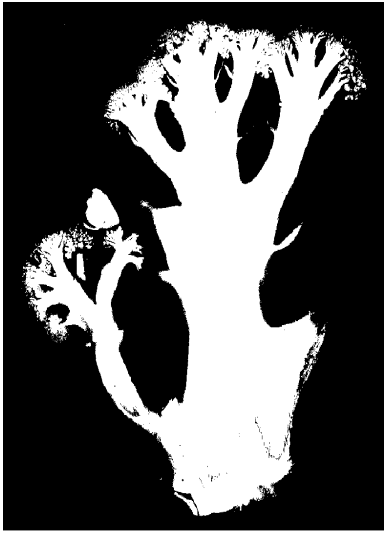
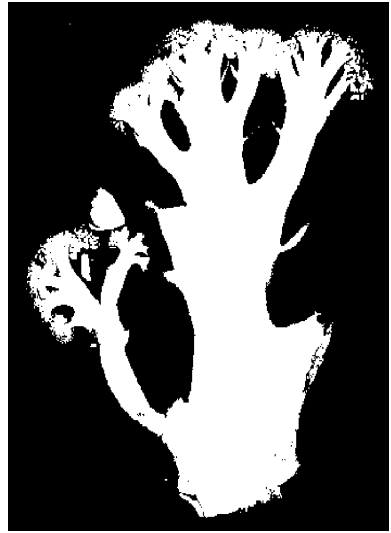
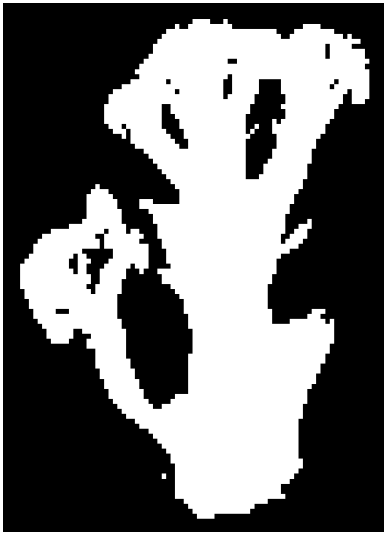
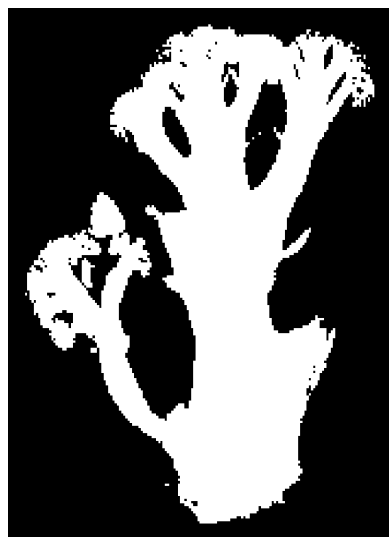
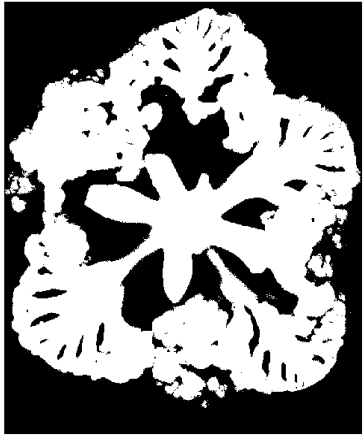
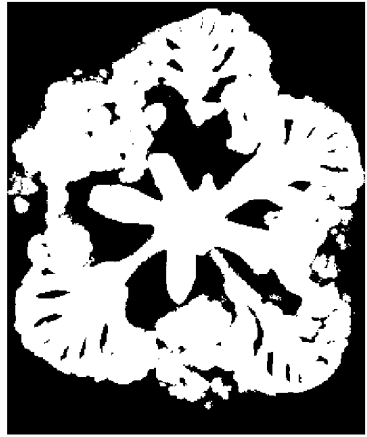
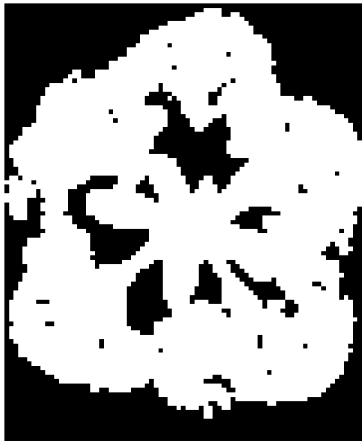
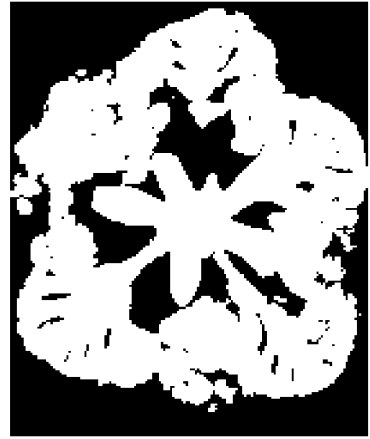
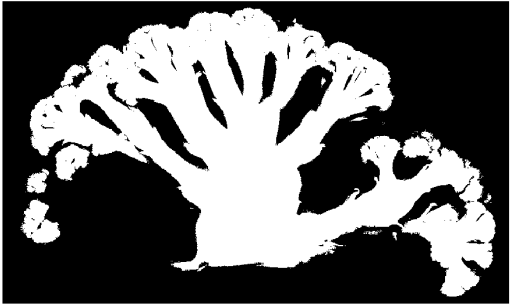
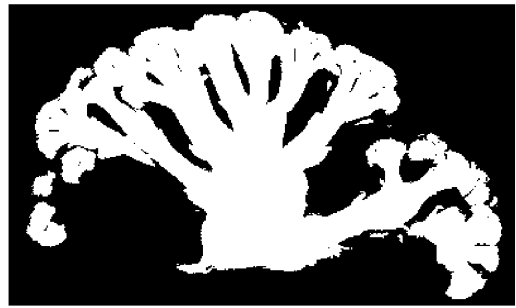
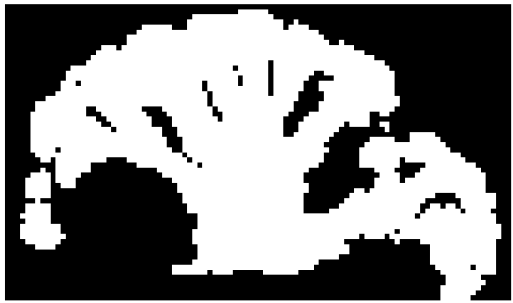
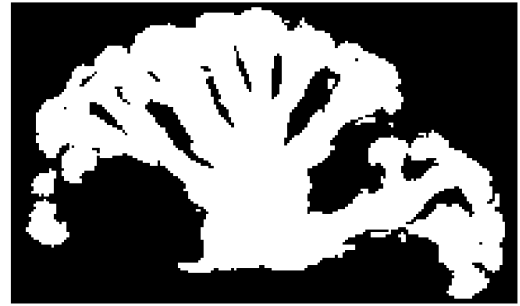
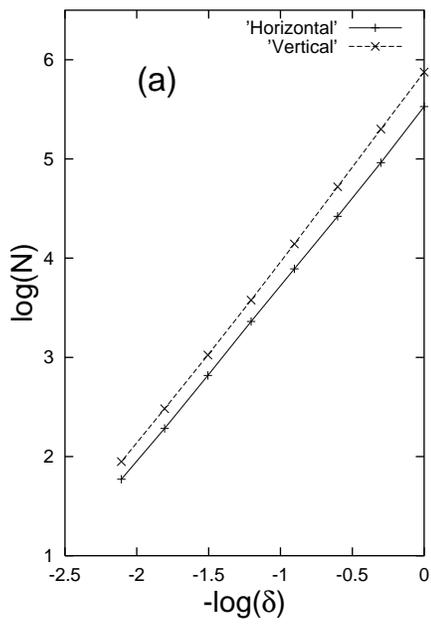
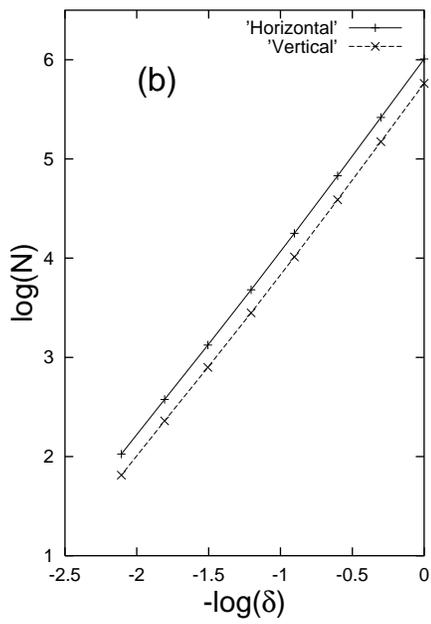
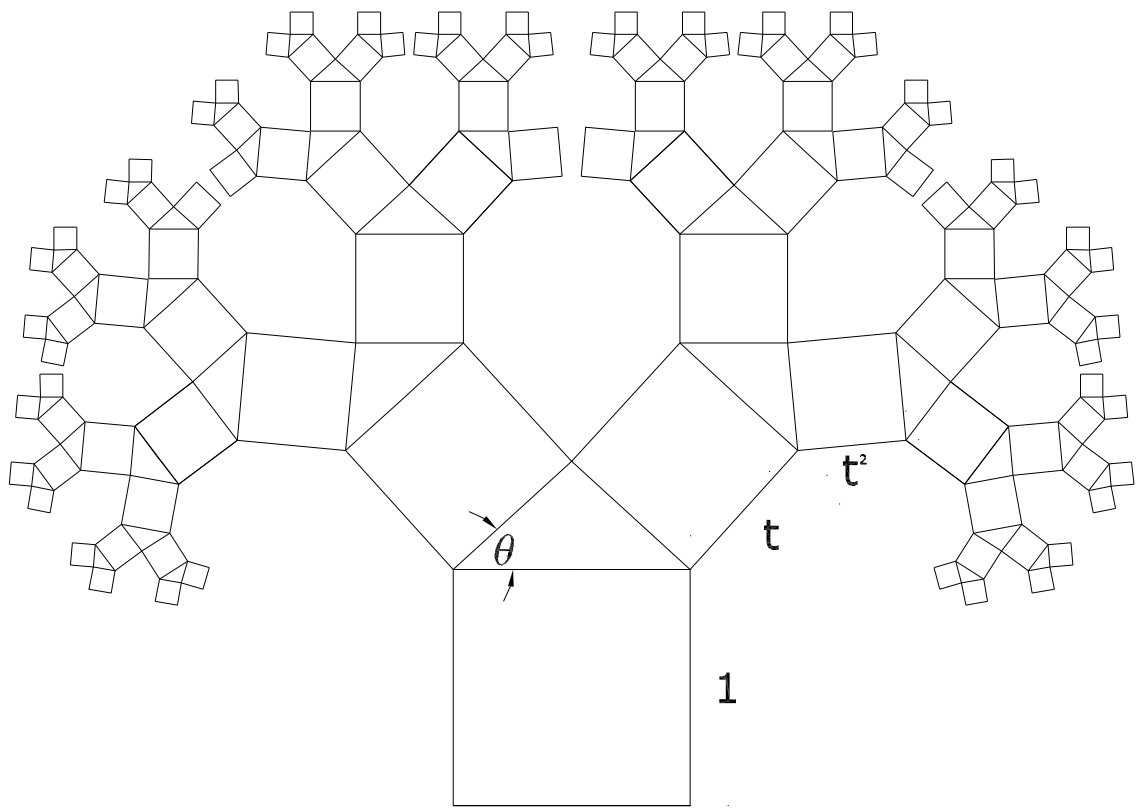
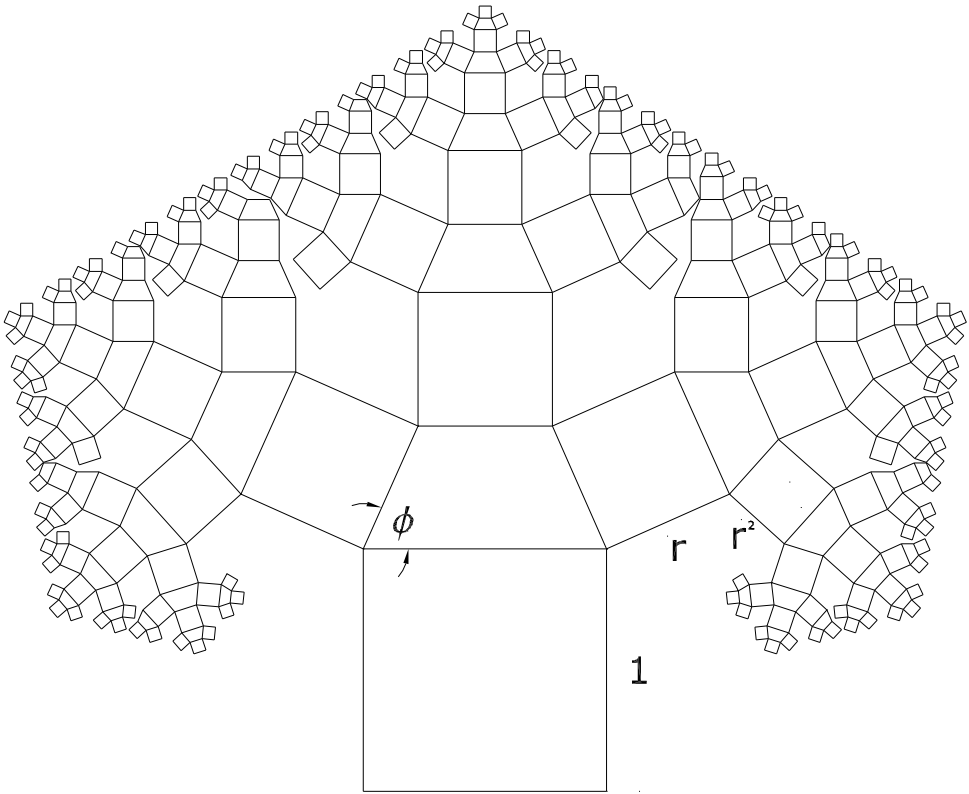
\end{document}